\documentclass[twocolumn,10pt,aps,prd]{revtex4-1}

\usepackage{amssymb,amsmath,amsthm}
\usepackage[pdftex]{graphicx} 
\usepackage{braket}
\usepackage[T1]{fontenc}
\usepackage[latin1]{inputenc}
\usepackage[english]{babel}

\usepackage{float}
\usepackage{braket}
\usepackage{slashed}
\usepackage{commath}
\usepackage{diagbox}
\usepackage{color}
\usepackage{bm}
\usepackage[breaklinks=true]{hyperref} 
\hypersetup{colorlinks=true,citecolor=blue,linkcolor=blue,urlcolor=blue}
\usepackage{booktabs}

\usepackage{tikz}
\usetikzlibrary{positioning,arrows,patterns}
\usetikzlibrary{decorations.markings} 
\usetikzlibrary{decorations.pathmorphing}

\addto\captionsenglish{}

\addto\captionsenglish{}

\renewcommand{\epsilon}{\varepsilon}
\newcommand{\mb}[0]{\mathbf}

\renewcommand{\rm}[1]{\mathrm{#1}}
\newcommand{\Ai}{\mathrm{Ai}}
\newcommand{\Bi}{\mathrm{Bi}}
\newcommand{\p}{\partial}
\newcommand{\w}{w^{\rm{add}}}
\newcommand{\wt}{w^{\rm{tot}}}
\newcommand{\Tg}{T_{\rm{grav}}}

\begin{document}
\title[]{Decay of a Massive Particle in a Stiff Matter Dominated Universe}

\author{Juho Lankinen}
\email{jumila@utu.fi}
\affiliation{Turku Centre for Quantum Physics, Department of Physics and Astronomy, University of Turku, 20014, Finland}

\author{Iiro Vilja}
\email{vilja@utu.fi}
\affiliation{Turku Centre for Quantum Physics, Department of Physics and Astronomy, University of Turku, 20014, Finland}

\begin{abstract}

In the presence of a gravitational field decay rates may significantly differ from flat space equivalent. By studying mutually interacting quantum fields the decay rates can be calculated on a given spacetime. This paper presents the calculation of the transition probability for the decay of a massive scalar particle in a stiff matter dominated universe. We find that due to the precence of a gravitational field a finite correction to the transition probability is added which depends inversely on the mass.  Moreover the decay rate is smaller and lifetime of the particles is longer compared to flat space. The mass dependence is such that the lifetime of lighter particles is prolonged more compared to heavier particles. This result may be of significance when studying cosmological situations involving stiff matter.
\end{abstract}

\maketitle

\section{Introduction}

Quantum field theory in curved spacetime, the study of propagating quantum fields on an unquantized background, is currently the best tool for investigating nature at its most fundamental level.  
One of the most striking features of this theory is the phenomenon of gravitational particle creation out of the vacuum. From the seminal works of Parker \citep{Parker:1968,Parker:1969,Parker:1971} through numerous present day studies considerable amount of work has been devoted to this aspect; see for example \citep{Birrell_Davies,Parker_Toms} and references therein. Although free fields are an important facet of study, realistic fields tend to interact with each other. Regarding this aspect, self-interacting quantum fields on an unquantized background have received a lot of attention, where investigations have mainly been confined to the problem of renormalization.  However, a closely related topic of mutually interacting fields, where the interest is on what effects the process of mutual interaction between two different quantum fields has on the particle production, has only scarcely been investigated in the literature. Few extensive studies were made some time ago by Audretsch and Spangehl \citep{Audretsch_Spangehl:1985,Audretsch_Spangehl:1986,Audretsch_Spangehl:1987}, Birrell, Davies and Ford \citep{Birrell_etal:1979} and Lotze \citep{Lotze:1989a,Lotze:1989b}. More recent studies have been concerned with QED processes in de Sitter spacetime \citep{Cotaescu:2013,Blaga:2015,Crucean:2015} and also in radiation dominated universe \citep{Tsaregorodtsev:2004}.{\color{red}{}}

Of physical interest in interacting field theories are the calculations of cross sections, decay rates and lifetimes. Due to the dynamical nature of the spacetime, these may differ greatly from those obtained from ordinary special relativistic quantum field theory which may be only of limited applicability in cosmological situations. It is therefore not only interesting but necessary to investigate the decay of particles in a given dynamical spacetime to understand under what conditions and to what extent the Minkowskian results are modified.
However, a major drawback usually found in cosmological models in the context of quantum field theory in curved spacetime is that they are not very realistic. One usually has to 'complete' the spacetime by introducing an in region at negative infinity in order to well define the quantum field modes, or to facilitate exact calculations an unrealistic interaction has to be introduced. In our previous article \citep{Lankinen:2017} a cosmological model for stiff matter dominated universe was introduced where the in region is not constructed at negative infinity but near the spacetime singularity making the model physically more interesting. Stiff matter filled universe is also particularly interesting from cosmological perspective since the inflation field, when dominated by its kinetic energy, behaves like stiff matter. Therefore the immediate time after inflation may well be described by a stiff matter era. Recently there has also been a growing interest in cosmological models exhibiting an early stiff matter era \citep{deHaro:2016a,deHaro:2016b,deHaro:2016c,Oliveira:2011,Chavanis:2015}. 

In this paper we study quantum field theory in a stiff matter dominated universe by investigating the mutual interaction between two different fields. We provide a detailed analysis of the transition probability for the decay of a massive particle in this spacetime and especially we use a realistic interaction and cosmological model to provide a physically motivated study. 
Calculation of the decay rate in curved space is more complicated than in flat space and a couple of different methods for the calculation exists. The method introduced in Refs. \citep{Boyanovsky_Holman:2011,Boyanovsky:2012} using Wigner-Weisskopf method an the one introduced in Ref. \citep{Audretsch_Spangehl:1985} which uses the method of added-up probabilities. We will use the latter and to use this concept we will use conformally coupled fields which are conformally coupled to gravity. We show that due to the dynamical background the decay rate obtains finite additive correction which depends inversely on the mass of the particle. 
Moreover the decay of a scalar particle in a stiff matter dominated universe is slower than in Minkowskian space.

This paper is structured as follows. Section \ref{sec:Preliminaries} contains the necessary background for dealing with particle decay. We give a brief review of the concept of added-up probabilities in Sec. \ref{sec:Added}. In Sec. \ref{sec:Decay} we obtain the transition amplitude and transition probability for the decay of a massive particle and discuss the obtained decay rate in Sec. \ref{sec:results}. Finally in Sec. \ref{sec:Discussion} we present the conclusions. We work in natural units $\hbar=c=1$ and the metric is chosen with a positive time component.

\section{Preliminaries}\label{sec:Preliminaries}
We consider the cosmological model presented in Ref. \citep{Lankinen:2017}. The universe is described by a four-dimensional spatially flat Friedmann-Robertson-Walker metric
\begin{align}
ds^2=a(\eta)^2(d\eta^2-d\bm{\mathrm{x}}^2)
\end{align}
given in conformal time $\eta$. By choosing $a(\eta)=b\eta^{1/2}$, $ \eta\in(0,\infty)$, this describes a stiff matter filled universe and the standard time scale factor $a(t)\propto t^{1/3}$.  Here the parameter $b$ controls the expansion rate of the spacetime. 
We consider a massive real scalar field $\phi$ and a massless scalar field $\psi$ propagating in this spacetime which are conformally coupled to gravity. The Klein-Gordon equation takes in this case the form 
\begin{align}\label{eq:KG}
(\square +m^2+R/6)\phi(\eta, \mb x)=0,
\end{align}
where $\square$ is the covariant d'Alembert operator and $R$ denotes the Ricci scalar. For massless field take $m=0$. 
The positive frequency plane-wave solutions of Eq.\eqref{eq:KG} in the asymptotic future are \citep{Lankinen:2017}
\begin{align}\label{eq:Stiff_exact_solution}
u_\mb{p}(\eta, \mb x)&=\frac{e^{i\pi/12}\sqrt{2\pi}}{(2\pi)^{3/2}z^{1/3}}\Ai\Big(e^{i\pi/3}\frac{p^2+z^2\eta}{z^{4/3}}\Big)\frac{e^{i \mb{p \cdot x}}}{a(\eta)},
\end{align}
where $z:=mb, p:=|\mb p|$ and $\Ai$ denotes the Airy function. These modes define the out-vacuum. For a massless field  the corresponding mode solutions are obtained straightforwardly from the flat space solutions:
\begin{align}\label{eq:Massless_mode_solution}
v_\mb{k}(\eta, \mb x)&=\frac{1}{(2\pi)^{3/2}a(\eta)}\frac{1}{\sqrt{2k}}e^{i\mb{k\cdot x}-ik\eta},
\end{align}
where $k:=k^0=|\mb k|$.
The usual expansion for the field can be written as
\begin{align}
\phi(x)&=\sum_{\mb p}\big(a_{\mb p}u_{\mb p}(x)+a_{\mb p}^\dagger u^*_{\mb p}(x)\big),\\
\psi(x)&=\sum_{\mb k}\big(b_{\mb k}v_{\mb k}(x)+b_{\mb k}^\dagger v^*_{\mb k}(x)\big),
\end{align}
where $a_{\mb p},a_{\mb p}^\dagger$ and $b_{\mb k},b_{\mb k}^\dagger$ are the annihilation and creation operators for the fields  $\phi$ and $\psi$ respectively satisfying the commutation relations $[a_{\mb p'},a_{\mb p}^\dagger]=\delta_{\mb {p p'}}$ and  $[b_{\mb k'},b_{\mb k}^\dagger]=\delta_{\mb {k k'}}$ with other commutators vanishing. The vacuum state $\ket{0}$ is defined through $a_{\mb p}\ket{0}=0, \forall\, \mb p$. For a  massless field with conformal coupling there is no particle creation due to expansion of spacetime and therefore the in and out vacuums agree with each other. For massive fields the in and out vacuums differ from each other.
 Many particle states are constructed in the usual way by acting repeatedly to the vacuum state by the creation operator.

We consider interacting field theories within the general prescription \citep{Birrell_Davies}. The interaction between the two fields is given by the Lagrangian
\begin{align}\nonumber
\mathcal{L}=&\frac{\sqrt{-g}}{2}\big\{\p_\mu \phi \p^\mu\phi-m^2\phi^2-\frac{R}{6}\phi^2+\p_\mu\psi\p^\mu \psi-\frac{R}{6}\psi^2\big\}\\
&+\mathcal{L_I},
\end{align}
where $g$ stands for the determinant of the metric.
For the interaction term we choose
\begin{align}\label{eq:L_I}
\mathcal{L}_I:=-\sqrt{-g}\lambda \phi \psi^2,\quad \lambda>0.
\end{align}
The $S$-matrix scheme is applied in the Furry picture with gravity acting as the external unquantized field and the $S$-matrix is given as
\begin{align}
S=\lim_{\alpha\to 0^+}\hat{T}\exp\Big(i\int\mathcal{L_I}e^{-\alpha\eta}d^4x\Big),
\end{align}
where $\hat{T}$ denotes the time-ordering operator. The exponential factor $e^{-\alpha\eta}$ acts as a switch-off for the interaction for large times with $\alpha$ being a positive constant and called the switch-off parameter.
The perturbative expansion of the $S$-matrix for the interaction \eqref{eq:L_I} gives
\begin{align}
S=1-i\lambda A+\mathcal{O}(\lambda^2)
\end{align}
with
\begin{align}\label{eq:A_integraali}
A:=\lim_{\alpha\to 0^+}\int \hat{T}\phi \psi^2 e^{-\alpha\eta}\sqrt{-g}\, d^4x.
\end{align}
We consider only tree level processes for which the transition amplitude is defined as 
\begin{align}
\mathcal{A}:= \braket{\rm{out}|A|\rm{in}}.
\end{align}

\section{Added-up probability}\label{sec:Added}
The interpretation of transition amplitudes and decay rates is more complicated in curved space than in flat space. One of the reasons is that in a dynamically expanding universe there is creation of particles out of the vacuum which interferes with the process of mutual interaction. A particle counter for massive particles always registers the combined effect of mutual interaction and background. One way around this problem is to consider a particle counter based on massless particles, since in a conformally flat spacetime conformally coupled massless particles are not influenced by the background spacetime. A registered massless particle has always been solely created or influenced by the mutual interaction. This is the key behind the concept of added-up probability introduced by Audretsch and Spangehl in Ref. \citep{Audretsch_Spangehl:1985}. The physically measurable quantity is then the probability that a certain massless state is found regardless of the states of the massive fields. This is achieved by summing over all massive states. Next we give a brief introduction to this method.

\begin{figure}[H]
\centering
\includegraphics[scale=1]{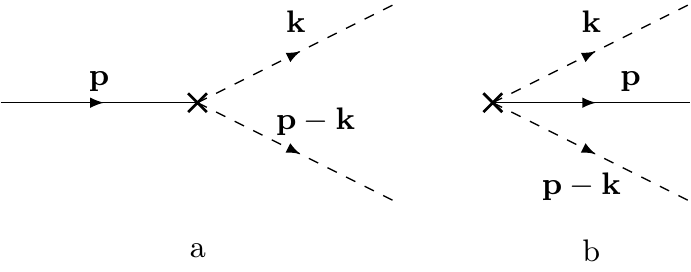}
\caption{Diagrams contributing to the added-up decay probability. The solid line corresponds to the massive particle and the dashed lines to massless particles.}\label{fig:1}
\end{figure}

Restricting to our case of processes of first order in  $\lambda$ with one ingoing massive mode with momentum $\mb p$ leading to two outgoing massless modes with momenta $\mb k_1$ and $\mb k_2$, the added-up probability is defined as \citep{Audretsch_Spangehl:1985}:
\begin{flalign}\nonumber
&\w_{\phi\rightarrow \psi\psi}(\mb p,\mb k_1,\mb k_2)\\
&=\lambda^2\sum\limits_{n=0}^\infty\frac{1}{n!}\sum\limits_{\mb{q}_1,...,\mb{q}_n}\lvert\braket{\rm{out}, 1^\psi_{\mb{q}_1},...,1^\psi_{\mb{q}_n}  1^\psi_{\mb{k}_1}1^\psi_{\mb{k}_2}|A|1^\phi_{\mb{p}}, \rm{out}}\rvert^2.
\end{flalign} 
Since both in and out states are complete, the added-up probability can be expressed with respect to either of these states. We choose the out state since for the model we are using the out state is an exact solution.
Audretsch and Spangehl argue, that restricting to those outgoing $\psi$ particle modes which fulfill the $3$-momentum conservation law $\mb k_1+\mb k_2=\mb p$, the added-up probability resembles closest to what one would like to attribute to a decay process. It contains the minimal admixture of creation processes from the vacuum which are indistinguishable from the decay process itself. With $A$ given in Eq. \eqref{eq:A_integraali}, the added-up probability now takes the form
\begin{align}\nonumber\label{eq:w_add}
\w_{\phi\rightarrow \psi\psi}(\mb p,\mb k,\mb{p-k})=&\lambda^2\Big\{ \lvert \braket{\rm{out}, 1^\psi_{\mb{k}}1^\psi_{\mb{p-k}}\lvert A \rvert 1^\phi_{\mb{p}}, \rm{out}}\rvert^2\\
&+\lvert\braket{\rm{out},1^\phi_{\mb{-p}} 1^\psi_{\mb{k}}1^\psi_{\mb{p-k}}\lvert A \rvert 0, \rm{out}}\rvert^2 \Big\},
\end{align}
where $\mb k_1=\mb k$ and $\mb k_2=\mb{p-k}$. The corresponding Feynman diagrams are given in Fig. \ref{fig:1}.
\noindent The first term in \eqref{eq:w_add} corresponds to diagram a of Fig. \ref{fig:1}, while the second term refers to diagram b. 
The total decay probability is obtained by summing over all the $k$-modes:
\begin{align}\label{eq:w_tot}
w^{\rm{tot}}_{\phi\rightarrow \psi\psi}=\sum_{\mb k} \w_{\phi\rightarrow \psi\psi} (\mb p,\mb k,\mb{p-k}).
\end{align}

\section{Decay of the Massive Particle}\label{sec:Decay}
Within the framework given in previous sections, we consider the $\phi\psi^2$ interaction illustrated by the Feynman diagrams in Fig. \ref{fig:1}. In the absence of a gravitational field, the massive $\phi$ particle decays into two massless $\psi$ particles as in diagram a. When a gravitational field is present the process shown in diagram b, where a massive and two massless particles are created simultaneously, can occur due to lack of energy conservation.
Next, we carry out the exact calculation of the transition amplitudes and the total probability.

\subsection{Transition amplitude}
Using the solutions \eqref{eq:Stiff_exact_solution} and \eqref{eq:Massless_mode_solution} together with Eq. \eqref{eq:A_integraali}, the transition amplitude of diagram a of Fig.\ref{fig:1} has the following form:
\begin{align}\nonumber
&-i\lambda\braket{\rm{out},1_{\mb{k}_1}^\psi 1_{\mb{k}_2}^\psi|A|1_{\mb{p}}^\phi,\rm{out}}\\\nonumber
=&\frac{-ib\lambda e^{i\pi/12}\sqrt{\pi}\delta(\mb{p}-\mb{k}_1-\mb{k}_2)}{\sqrt{2k_1k_2}z^{1/3}(2\pi)^{3/2}}\\
&\times \lim_{\alpha\to 0^+}\int_0^\infty \!\! \Ai\Big(e^{i\pi/3}\frac{p^2+z^2\eta}{z^{4/3}}\Big)e^{[i(k_1+k_2)-\alpha]\eta}\sqrt{\eta}\,d\eta.
\end{align}
The delta function expresses the conservation of momentum among the created particles. The stiff matter expansion law allows a particle definition for the in-region which is at $\eta\rightarrow 0$ and for the out-region at $\eta\to\infty$ so we take the conformal time integral to go from zero to infinity. 
Changing to a dimensionless variable $u=(p^2+z^2\eta)/z^{4/3}$ we obtain
\begin{align}\nonumber
&-i\lambda\braket{\rm{out},1_{\mb{k}_1}^\psi 1_{\mb{k}_2}^\psi|A|1_{\mb{p}}^\phi,\rm{out}}\\
&=\frac{-ib\lambda e^{i\pi/12}\sqrt{\pi}\delta(\mb{p-k_1-k_2})}{\sqrt{2k_1k_2}(2\pi)^{3/2}z^2}I_A(u,k_1+k_2),
\end{align}
where
\begin{align}\nonumber
I_A(u,k_1+k_2)
=&\lim_{\alpha\to 0^+}\!\int_{p^2/z^{4/3}}^\infty \sqrt{z^{4/3}u-p^2}\Ai(e^{i\pi/3}u)\\
&\times e^{[i(k_1+k_2)-\alpha](z^{4/3}u-p^2)/z^2}du.
\end{align}
The amplitude of diagram b has the same factors as in the amplitude of diagram a, but with $\delta(\mb p-\mb k_2-\mb k_2)$ replaced by $\delta(\mb q+\mb k_1+\mb k_2)$. Furthermore the exponential in the argument of the Airy function is replaced by a negative sign. The amplitude of diagram b is given by
\begin{align}\nonumber
&-i\lambda\braket{\rm{out},1_{\mb{q}}^\phi 1_{\mb{k}_1}^\psi 1_{\mb{k}_2}^\psi|A|0,\rm{out}}\\
&=\frac{-ib\lambda e^{i\pi/12}\sqrt{\pi}\delta(\mb{q+k_1+k_2})}{\sqrt{2k_1 k_2}(2\pi)^{3/2}z^2}I_B(u,k_1+k_2),
\end{align}
where
\begin{align}\nonumber
I_B(u,k_2+k_2)
=&\lim_{\alpha\to 0^+}\!\int_{q^2/z^{4/3}}^\infty  \sqrt{z^{4/3}u-q^2}\Ai(e^{-i\pi/3}u)\\
&\times e^{[i(k_1+k_2)-\alpha](z^{4/3}u-q^2)/z^2}du.
\end{align}
Again, the outgoing particle modes are constrained to those which fulfill the $3$-momentum conservation law.
With these two amplitudes the total added-up probability can now be calculated.

\subsection{Transition probability} 
We want to calculate the total transition probability $\wt$ as defined in \eqref{eq:w_tot}. In order to allow an exact calculation, we perform integration for the case of decaying particles at rest ($\mb p=0$) obtaining
\begin{align}\label{eq:TotalProbability}
\wt_{\phi\rightarrow\psi\psi}&=\frac{\lambda^2}{48m^2}\Big( T_\eta-\frac{\sqrt{3}}{3\pi} \Big).
\end{align}
The details of this calculation are given in the appendix. The infinite duration of the interaction is contained in the term $T_\eta$ and its explicit form is given by
\begin{align}\nonumber\label{eq:AiryBairy}
T_\eta:=&T^2\text{Ai}(-T)^2+ T^2\text{Bi}(-T)^2+ T\text{Ai}'(-T)^2\\
&+T\text{Bi}'(-T)^2+\Bi(-T)\Bi'(-T)+\Ai(-T)\Ai'(-T),
\end{align}
where $T$ denotes the scaled conformal time cutoff $T=z^{2/3}\eta$. The total probability contains an infinite part and an additive finite part which is the contribution of the gravitational field. We stress that Eq. \eqref{eq:TotalProbability} is an exact result and that in the limit of $T\to 0$ it goes to zero, as it should.
To better understand the behavior of \eqref{eq:TotalProbability}, we look at its asymptotic values for large $T$. Taking only the leading terms in the asymptotic series we have
\begin{align}
T_\eta\sim\frac{2}{\pi}T^{3/2}=\frac{2}{\pi}m b\eta^{3/2},
\end{align}
where we have restored $z=mb$. In conformal time the leading two terms in the asymptotic expansion of the decay probability is given as
\begin{align}\label{eq:End_Result_Conformal}
\wt_{\phi\rightarrow\psi\psi}\sim \frac{\lambda^2 b}{24\pi m}\Big(\eta^{3/2}-\frac{1}{2\sqrt{3}\,bm} \Big).
\end{align}
The conformal time variable $\eta$ can be changed back to the coordinate time $t$ by using the relation
\begin{align}
t=\frac{2}{3}b \eta^{3/2}.
\end{align}
Asymptotically the total transition probability can be expressed in standard coordinate time as
\begin{align}\label{eq:End_result}
\wt_{\phi\rightarrow\psi\psi}\sim \frac{\lambda^2}{16\pi m}\Big( t-\frac{1}{3\sqrt{3}\,m} \Big).
\end{align}
This is only valid when $t$ is large enough.

\section{The decay rate}\label{sec:results}

In flat spacetime the infinite time would be divided out in order to obtain the reciprocal of lifetime, the decay rate. Due to the structure of \eqref{eq:End_result} this procedure is not so straightforward. To proceed, the decay rate can be considered as in Ref. \citep{Audretsch_Spangehl:1985} where one divides the finite additive part by a finite gravitational time $\Tg$ representing the duration of the gravitational influence. This can be defined as $\Tg=t_f-t_i$, where $t_i$ denotes the time when the gravitational field begins its influence and $t_f$ its end. The mean decay rate is then
\begin{align}\label{eq:DecayRate}
\Gamma_{\phi\rightarrow\psi\psi}=\frac{\lambda^2}{16\pi m}\Big(1-\frac{1}{3\sqrt{3}\,m\Tg} \Big).
\end{align}
The minus sign contained in the additive part can make the whole decay rate negative so, in order for the decay rate to make sense, we require it to be positive. This imposes the following restriction for the gravitational influence:
\begin{align}\label{eq:T_rajoite}
\Tg>\frac{1}{3\sqrt{3} m}.
\end{align}
The differential decay rate is obtained straightforwardly from Eq. \eqref{eq:End_result} and it is given by
\begin{align}\label{eq:Differential_decay}
\frac{d \wt}{dt}=\frac{\lambda^2}{16\pi m}.
\end{align}

The result \eqref{eq:DecayRate} can be compared with the flat space equivalent obtained by using the same method of calculation. Using a normal plane wave solutions, the Minkowskian result for the decay rate using added-up probabilities, with $\mb p=0$, is
\begin{align}
\Gamma^{\rm{Mink}}_{\phi \rightarrow  \psi \psi}=\frac{\lambda^2}{16\pi m}.
\end{align}
It should be noted that in this calculation only diagram a of Fig. \ref{fig:1} is taken into account since the second diagram does not contribute in flat space. Comparing the Minkowskian and mean decay rates the following observations can be made.

The minus sign in Eq. \eqref{eq:DecayRate} implies that the contribution of the gravitational field decreases the decay rate prolonging the lifetime of the particles. Since the correction term is inversely proportional to the mass of the particle in question, this effect is most notable for light scalars. The appearance of the minus sign also imposes a restriction to the gravitational influence in order to keep the decay rate positive. It must be stressed however that the result \eqref{eq:DecayRate} is the first order asymptotic expansion which means that it is only valid if $t$ is sufficiently large. This in turn implies that $T_{\rm{grav}}$ is large already. As the time gets smaller, next to leading order terms have to be taken into account since then they start to contribute. Hence, the restriction \eqref{eq:T_rajoite} is the minimum requirement that the decay rate with only leading order terms is positive. The exact decay rate remains always positive. This brings us to the question on how significant this relative correction term actually is in a real setting. Since the formula for the decay rate \eqref{eq:DecayRate} is only valid under the restriction of Eq. \eqref{eq:T_rajoite} and $T_{\rm{grav}}$ is usually much longer than inverse of mass the relative correction term is in practice quite small. Considering a practical setting, it thus seems that the difference in decay rates between an expanding universe and the Minkowski space is not very significant. However it cannot be said that the relative correction term can be neglegted altogether, in particular for $t\sim m$ when the full equation \eqref{eq:AiryBairy} should be used.

Since only tree level processes are considered, the higher order perturbative corrections affect the results as well as other processes in a physically realistic model. In particular, there may be kinematically forbidden flat space processes, e.g., decay via trilinear self-interaction, which may contribute to the rate. These processes are however much more challenging tasks \citep{Boyanovsky_Holman:2011}.

A feature that must be addressed is that the metric used here does not have a well defined Minkowskian limit as $b$ tends to zero.
Therefore the comparison of the obtained decay rate \eqref{eq:DecayRate} to the Minkowskian counterpart should be taken with some caution.  One could of course introduce a metric of the form $a(\eta)^2=d^2+b^2\eta$ with $d$ being a constant which could be set equal to one. This way the metric would have a well defined Minkowskian limit as $b\to 0$, but it turns out that the correction term does not. The cause of this can be traced back to the conformal time integral and the effect of taking the $b\to 0$ limit causes the lower limit to extend to infinity. One notable difference between our result and that of Ref. \citep{Audretsch_Spangehl:1985} is the missing factor of one half in front of the Minkowskian term in our result. The appearance of $1/2$ in Ref. \citep{Audretsch_Spangehl:1985} was explained to be the result of using an interaction term which was divided by the scale factor. The use of a realistic coupling resulting in no additional terms in the Minkowskian term could be a general feature.

We have also limited our study to a universe filled with stiff matter. This choice was motivated not only by cosmological interest but also by practical matters, since the field modes for a universe which is dominated by stiff matter at all times are known. Although processes occurring in radiation or matter dominated phase are evidently relevant, these situations prove to be much more difficult since the field modes for a massive particle in a matter dominated phase are not known and are only known for a universe which is asymptotically radiation dominated \citep{Audretsch_Schafer:1978}.

Finally we may speculate about why the decay rate has an explicit leading Minkowskian part in $T_{\rm{grav}}^{-1}$ expansion both here and in Ref. \citep{Audretsch_Spangehl:1985}. General coordinate invariance allows to set up a quasi-Minkowskian coordinate system to each point in the path of a single decaying particle. For the decay rate calculation one constructs a wave packet which is peaked around the momentum $\mb p$.  Hence in the vicinity of the particle path the mode solutions are quasi-Minkowskian and the decay rate obtains a Minkowskian contribution nearby the path. However, the particle is not completely localized in spacetime and therefore when the whole wave packet along the path is taken into account there is a curvature induced addition. More rigorous treatment is surely needed to establish this hypothesis on a firmer basis or disprove it.

\section{Conclusions}\label{sec:Discussion}

The presence of a gravitational field may significantly alter the decay rates in a given dynamical background and the choice of the scale factor affects greatly on how the contribution of the gravitation field increases or decreases the decay rate. 
Our results imply that in a stiff matter dominated universe the Minkowskian results are applicable when considering very massive particles, but with light particles there is a gravitational correction term which should be taken into account.
This result can be of significance when studying cosmological situations although it is not entirely clear, due to the construction of the added-up probability, how the decay rate in itself should be used e.g., in the Boltzmann equations. Cosmological situations involving a stiff matter era include for example reheating scenarios which could be a point of study for future research.
\begin{acknowledgments}
J.L would like to acknowledge the financial support from the University of Turku Graduate School (UTUGS).
\end{acknowledgments}

\appendix*
\section{Total probability}\label{sec:appendix}
To calculate the total probability \eqref{eq:w_tot} we pass to continuum normalization and perform the $\mb k_2$ integration:
\begin{flalign}\nonumber
w^{tot}\!&=\frac{\lambda^2b^2\pi}{2(2\pi)^{3/2}}\int_{-\infty}^\infty \frac{d^3\mb k}{k|\mb{p-k}|} \Big|\!\lim_{\alpha \to 0^+}\!\int_{p^2/z^{4/3}}^\infty\!\!\!  \sqrt{z^{4/3}u-p^2}\\
&\times e^{(i(k+|\mb{p-k}|)-\alpha)(z^{4/3}u-p^2)/z^2}\Ai(e^{i\pi/3 }u) du \Big|^2.
\end{flalign}
In obtaining this we have used the fact that  $\lvert I_A(u, k+|\mb{p-k}|)\rvert^2=\lvert I_B(u,- k-|\mb{p-k}|)\rvert^2$.
Next we are going into the rest frame of the incoming particle $\mb p=0$ and changing the $k$-integration variable to $k'=2kz^{-2/3}$. Using spherical coordinates we obtain

\begin{align}\label{eq:w_tot_exact}
\wt&=\frac{\lambda^2b^2}{8\pi z^2}I,
\end{align}
where
\begin{align}\label{eq:I_integraali}
I=\int_{-\infty}^\infty dk'\Big|\lim_{\alpha' \to 0^+}\int_0^\infty\!\!\! \sqrt{u}\,e^{(ik'-\alpha')u}\Ai(e^{i\pi/3 }u) du \Big|^2.
\end{align}
Here $\alpha'=\alpha/z^{2/3}$. In order to calculate \eqref{eq:I_integraali} we notice that it is essentially a three dimensional integral containing the $k'$-integration and a two dimensional integral from the absolute value. The $k'$-integration can be treated as a distribution, since contains an integral of the form 
\begin{align}\label{eq:delta_function}
\int_{-\infty}^\infty e^{i k'(x-y)}dk'=2\pi\delta(x-y).
\end{align}
With the Eq. \eqref{eq:delta_function} two of the integrations in  \eqref{eq:I_integraali} can be performed leaving the integral
\begin{align}
I=\frac{\pi}{2}\lim_{\alpha'\to 0}\int_0^\infty e^{-2u\alpha'}u[\Ai(-u)^2+\Bi(-u)^2]du.
\end{align}
The Airy functions of the integrand can be written in the following integral form \citep{Reid:1995}:
\begin{align}
\Ai(-u)^2+\Bi(-u)^2=\int_0^\infty\frac{e^{-t^3/12-u t}}{\pi^{3/2}\sqrt{t}}dt.
\end{align}
With the help of this integral form we are left with the double integral
\begin{align}\label{eq:I_alpha}
I=\frac{\pi}{2\pi^{3/2}}\lim_{\alpha'\to 0}\int_0^\infty\int_0^\infty \frac{ue^{-t^3/12-ut-2\alpha'u}}{\sqrt{t}}dtdu.
\end{align}
The $u$-integration, which is essentially the time variable, diverges so we introduce a cutoff at $u=T$. Changing the order of integration and taking the $\alpha'$ limit, we get
\begin{align}\nonumber
I_T&=\int_0^\infty \frac{e^{-\frac{t^3}{12}-t T} \left(-t T+e^{t T}-1\right)}{2\sqrt{\pi}t^{5/2}}dt\\
&=\frac{\pi}{6}\Big(T_\eta-\frac{\sqrt{3}}{3\pi}\Big),
\end{align}
where
\begin{align}\label{eq:Airy1}\nonumber
T_\eta:=&T^2\text{Ai}(-T)^2+ T^2\text{Bi}(-T)^2+ T\text{Ai}'(-T)^2\\
&+T\text{Bi}'(-T)^2+\Bi(-T)\Bi'(-T)+\Ai(-T)\Ai'(-T)
\end{align}
Hence, the total added-up probability reads as
\begin{align}
\wt_{\phi\rightarrow\psi\psi}&=\frac{\lambda^2}{32m^2}\Big( T_\eta-\frac{\sqrt{3}}{3\pi} \Big).
\end{align}

\end{document}